\documentclass[amsmath,amssymb,floatfix,lengthcheck,superscriptaddress]{revtex4-1}
\usepackage{newtxtext}
\usepackage{graphicx}
\usepackage{dcolumn}
\usepackage{bm}
\usepackage{hyperref}
\hypersetup{colorlinks, 
    linkcolor={blue!75!black!80!yellow},
    citecolor={blue!75!black!80!yellow}, 
    urlcolor={blue!75!black!80!yellow}
    }
\usepackage{xcolor} 
\usepackage{soul}
\usepackage{bibunits}

\defaultbibliographystyle{apsrev4-1}

\begin{document}

\title{Native point defects from stoichiometry-linked chemical potentials in cubic boron arsenide}

\author{Yaxian Wang}
\author{Wolfgang Windl}%
 \email{windl.1@osu.edu}
\affiliation{Department of Materials Science and Engineering, The Ohio State University, Columbus, OH 43210
}%

\date{\today}

\begin{abstract}
The presence of a point defect typically breaks the stoichiometry in a semiconductor. For example, a vacancy on an A-site in an AB compound makes the crystal B-rich. As the stoichiometry changes, so do the chemical potentials.
While the prevalent first-principles methods have provided significant insight into characters of point defects in a transparent manner, the crucial connection between crystal stoichiometry and chemical potentials is usually not made.
\textcolor{black}{However, ad hoc choices for chemical potentials can lead to nonphysical negative formation energies} in some \textcolor{black}{Fermi level} ranges, along with questions about charge balance.
Herein, we \textcolor{black}{formulate a canonical} framework \textcolor{black}{describing how} the chemical potential of each element is directly linked to the \textcolor{black}{composition} of the crystal \textcolor{black}{under (off-)stoichiometric conditions }instead of the ad hoc assumption that the chemical potential is the elemental limit under a certain growth condition.
Consequently, the chemical potential changes with the \textcolor{black}{Fermi level} within the band gap, and the formation energies \textcolor{black}{are positive}. 
Using such an approach, we present \textit{ab initio} results for native point defects in BAs, a semiconductor with ultra-high room temperature thermal conductivity.
\textcolor{black}{We find that antisites are the constitutional defects in off-stoichiometric material, 
while B$_\mathrm{As}$ antisites and B vacancies dominate in the stoichiometric material.}
We further discuss the thermodynamic equilibrium and charge neutrality point in BAs \textcolor{black}{in light of} our stoichiometry-determined chemical potentials.
As discussed, our work offers a more applicable and accessible approach to tackle defect formation energies in semiconductors, especially the ones with wide gap where negative formation energies are commonly seen.
\end{abstract}

\maketitle
\begin{bibunit}

\section{\label{Sec:intro}Introduction}
Point defects are the only \textcolor{black}{thermodynamically stable crystal defects at finite temperatures.}
For functional, optical, and electronic materials, point defects influence a material's properties arguably more than any other factor. 
While defects or impurities can serve as dopants to achieve either $p$- or $n$-type conduction, they are mostly detrimental to the precise control of optical and transport properties.
As a result, extensive studies from both experiment and theory have been put forward to better understand the nature of both native point defects and impurities, and thus improve control of their formation in semiconductors.
While experimental observation of point defects is especially difficult and typically limited to indirect probing,\cite{JOHNSON201717,stoddard_duscher_windl_rozgonyi_2004,Asel18} their study has been one of the central points of computational materials science.
Especially first principles, or \emph{ab initio} calculations have been shown to be a powerful tool to identify the characteristics of point defects in functional materials, starting in the last two decades of the previous century such as reported in Refs.~\onlinecite{Scheffler88,laks1991role,Neuge94,windl1999first} to pick a few. 
Several review chapters and articles have also been published, summarizing a gamut of topics from the calculation of point defect energetics and dynamics to their combination with process simulations.\cite{Estreicher00,Windl2004diffusion,Windl2006,Freysoldt2014first} 

Generally speaking, based on DFT total energies for defective and perfect cells, the defect formation energies can be calculated using (for the example of an A vacancy in a stoichiometric AB compound, V$_\mathrm{A}$)
\begin{equation}
{\color{black}
E_f(\mathrm{V_A}) = E\mathrm{(V_A)}-E_{\rm perf}+\mu_{\rm A}+Q(E_F+E_{\mathrm{v}})},  
\label{eq:formation_energy}
\end{equation}
where \textit{Q} is the charge state, $E_F$ the \textcolor{black}{Fermi level} with $0<E_F<E_{\rm gap}$, $E_{\rm v}$ the energy of the top of the valence band, $\mu_{\rm A}$ the chemical potential of species A,
\textcolor{black}{and $E_{\rm perf}$ the total energy of the perfect supercell.} 
$E\mathrm{(V_{A})}$  is the total energy of a relaxed AB supercell with A vacancy. The same framework would hold if instead of total energies free energies were used.
Although the DFT the energies in Eq.~\ref{eq:formation_energy} can be corrected with a plenitude of schemes for e.g.\ finite size effects on energies\cite{Windl2004diffusion} and shallow dopant levels,\cite{Windl1998first} artifacts from the density functional used,\cite{Rampi12} charge interactions between periodic images,\cite{Makov95} and other artifacts,\cite{Freysoldt2014first} the understanding of the chemical potentials in Eq.~\ref{eq:formation_energy} has yet to be well converged.

The currently prevalent heat of formation (HoF) method focuses on the ultimate limits where one of the chemical potentials is aligned with the elemental solids.\cite{Freysoldt2014first}  
While this method provides a straightforward way to determine  defect formation energies, it does not explore the intrinsic connection between the stoichiometry and the chemical potentials because of the binary choice of chemical potentials along with the intermediate and often unconsidered thermodynamic ensemble situated between grand canonical and canonical.\cite{Freysoldt2014first}  
Without proper understanding, the HoF approach \textcolor{black}{may have pitfalls that lead to} defect energies and concentrations in unphysical limits, \textcolor{black}{may leave} charge neutrality typically unexplored, \textcolor{black}{and may lead} to focusing on the wrong prevalent point defects.

\textcolor{black}{An} aspect within work based on the HoF approach \textcolor{black}{that can be easily overlooked} is the connection between the Fermi level and charge neutrality point (CNP).
\textcolor{black}{Although explicitly defined recently in a review\cite{Freysoldt2014first} after decades of HoF use, few publications based on HoF discuss this aspect, which when neglected makes it difficult to interpret the traditional HoF results in a physical way in a gapped material.
More specifically, in} HoF, which is usually explored at zero temperature for a given chemical potential, the slope in the formation-energy-versus-Fermi-level plots corresponds to the point defect's charge.
\textcolor{black}{At zero temperature, charge neutrality,} \textcolor{black}{a fundamental requirement for point defects in any material with a band gap,\cite{Callister}} \textcolor{black}{can be identified by finding} \textcolor{black}{the lowest-energy intersection point between the line for a point defect with positive slope and the line for a point defect with negative slope with the same absolute values of slope.}
\textcolor{black}{There are no previous studies that address the CNP question from the HoF results, including }\textcolor{black}{discussing them} \textcolor{black}{in light of finite temperatures.}
When only considering the CNP, it can be several~eV away from mid gap in large-gap materials such as Al$_2$O$_3$ as a prototypical example~\cite{Choi13}, while the defect formation energies are so high that the defect concentrations will not shift the Fermi level significantly from mid gap in the material (for details see SM Sec.~S.1). When instead focusing on mid gap, the questions about charge neutrality and resulting electron and hole concentrations remain unanswered.

While thus the presently prevailing HoF method, whose approach was originally developed for compounds in contact with elemental solids,\cite{Scheffler88} focuses on the elemental limits of the constituent elements, the approach prposed by Zhang and Northrup in 1991\cite{Northrup91}  focuses on the difference between the chemical potentials $\Delta \mu=\mu_{\rm A}- \mu_{\rm B}$ under the requirement that the AB compound is stable with respect to the elemental chemical potentials, with no explicit values for $\mu_{\rm A}$ and $\mu_{\rm B}$ assumed in the input or received in the output. 
This allows to connect the defect concentrations to the overall stoichiometry through a quantitative link to chemical potential and Fermi level and thus eliminates the above discussed problems.
However, the somewhat reverse nature and the fact that the chemical potentials of the constituent elements are never explicitly determined may have kept it from being the standard approached used today.
\textcolor{black}{In addition, the formulation in terms of $\Delta \mu$ does not allow a straightforward extension to systems with more than two components.}
Here we will take this approach to a new level, where the desired stoichiometry is used as an input and where the explicit chemical potentials for all constituents can be determined. 

As demonstration system, we choose cubic BAs, where we study  the structure and energetics of its native point defects.
BAs was initially predicted to be the best thermal conductor with an ultrahigh lattice room-temperature thermal conductivity of 2000~W/(mK) in 2013~\cite{lindsay2013first}, and a few years later new theoretical work suggested a room temperature value of  1400~W/(mK)~\cite{lindsay17}.
Meanwhile, a number of groups have succeeded in synthesizing boron arsenide crystals of increasingly better quality~\cite{li2018high,tian2018unusual,kang2018experimental,tian2018seeded,kim2016thermal} and high thermal conductivities around 1300~W/(mK) at room temperature have been found. 
Although by now experiment and theory converge on a value for the thermal conductivity limit in high-quality crystals, the question about the most prevalent native point defects in BAs is still unsettled. 
Several theoretical attempts have been made to unravel the nature of the dominant native point defects, using their connection to off-stoichiometry and their detrimental effect on the thermal conductivity measured in BAs~\cite{lv2015experimental,protik2016ab,zheng2018antisite}. 
However, there has been no convergence on the constitutional point defects to date, i.e.\ the point defects responsible for off-stoichiometric compositions.\cite{mishra2012native} 
Initial work has focused on vacancies, which were chosen in an ad hoc fashion,\cite{lv2015experimental,protik2016ab} and subsequently, Ref.~\onlinecite{zheng2018antisite} suggested that B$_\mathrm{As}$-As$_\mathrm{B}$ antisite pairs should be the lowest-energy defects.
\textcolor{black}{However, the stoichiometric composition of the pair, which in Fig.~2 from Ref.~\onlinecite{zheng2018antisite} is the lowest-energy defect, violates the basic assumption of As-richness and indicates a flaw in the calculation: by adding the antisite pair as the {\it only} point-defect cluster out of the full ensemble of paired point defects, the antisite pair is not competing with any other cluster. 
Nevertheless, point defects very frequently have the tendency to cluster to minimize the number of mis-bonds, which gives the clusters lower energy than the sum of their constituent point defects and sometimes even lower than single point defects. 
The mismatch between the stoichiometry of the antisite pair and the stoichiometry premise thus can be taken as an indication  that if a more complete set of clusters would have been considered, there should be another cluster that is As-rich which has a lower energy at the CNP. 
Taking cues from other tetrahedral compound semiconductors, point defect complexes confirmed by experiments are indeed often more complex than a simple antisite pair, such as the V$_{\rm C}$C$_{\rm Si}$-Si$_{\rm C}$C$_{\rm Si}$ and Si$_{\rm C}$(C$_{\rm Si})_2$ complexes identified by EPR as predominant annealing products in SiC in Ref.~\citenum{pinheiro2004silicon}, which, translated to BAs, could be compatible with both the TEM images and the stoichiometry assumptions, while the predicted antisite pair from Ref.~\citenum{zheng2018antisite} contradicts the underlying stoichiometry.}
This discrepancy suggests a methodological deficiency after the initial assumption of As-rich chemical potentials.
Therefore, a theoretically sound answer to the question about the energetics of point defects in BAs is still outstanding, especially with a comprehensive examination of their connection to stoichiometry, which we will provide here. 
Since our goal is to have a method that shows better consistency between assumptions and results regarding chemical potentials and stoichiometry, we will restrict our examination to the case of single point defects, due to the \textcolor{black}{potential} misinterpretations resulting from comparing incomplete sets of defect complexes such as clusters to single point defects \textcolor{black}{ as we just discussed}.

In this paper, we extend the method proposed in Ref.~\citenum{Northrup91} and \textcolor{black}{reformulate a full canonical approach to} explicitly calculate the chemical potentials self-consistently at both 0~K and finite temperatures.
\textcolor{black}{By using the stoichiometry as input variable and allowing the chemical potentials to be Fermi-level dependent, we treat both stoichiometric and off-stoichiometric conditions within the same framework. We also show that the off-stoichiometric limit saturates to  ``constitutional point defects'' that have been defined within the \textcolor{black}{grand-thermodynamic-potential approach} developed within the field of intermetallics and alloys towards the end of the previous century\cite{Mayer95,foiles_daw_1987,Schott97,mishra2012native,hagen1998point}.}
We discuss how \textcolor{black}{our approach can be used to determine} physical chemical potentials, carrier concentrations and charge neutrality points in gapped materials.

We then present a systematic study of native point defects in cubic BAs, including vacancies, antisites and interstitials. 
Our results include the fully relaxed structures, their minimum-energy charge states, and their formation energies for the B-rich, As-rich and stoichiometric cases at 0~K and finite temperatures. 
For the latter, we calculate the fully temperature dependent free energy of formation within the quasi-harmonic approximation in the dilute concentration limit where the configurational entropy contribution is shown to be much smaller than that of vibrational entropy, and thus negligible in altering the equilibrium point defect concentrations.  
Our results at finite temperatures show that it is the boron vacancy and both antisites that play the most important role. 
In addition, we give a detailed discussion of CNP and thermodynamic equilibrium in BAs in light of these calculations, and  \textcolor{black}{determine quantitatively the $p$-type charge carrier concentration induced from self-doping in the frequently observed B-rich samples.}
These together validate our predictions as well as the underlying methodology.

\section{\label{sec:method}Methods}
\subsection{\label{sec:theoryapproach}Theoretical framework}
Based on the DFT total energies for defective and perfect cells as calculated in Sec.~\ref{sec:computational}, the defect formation energies can be calculated using Eq.~\ref{eq:formation_energy}.
Moreover, we account for the vibrational contribution to the free energies of formation at finite temperatures, $F_f(T)$, which we obtain using the quasi-harmonic approximation~\cite{baroni2001phonons} based on supercell $\Gamma$-point calculations as described in previous work~\cite{luo2009first} and summarized in Sec.~S2 of the Supplementary Material (SM). 
Free energies of formation of a certain point defect can be determined in analogy to Eq.~\ref{eq:formation_energy} by replacing $E$ ($E_f$) with $F$ ($F_f$).

For determining the necessary chemical potentials from Eq.~\ref{eq:formation_energy}, we propose \textcolor{black}{a canonical} approach where they are determined by the stoichiometry of the crystal, starting from the method discussed in Ref.~\citenum{Northrup91}.
Here, the two constraints necessary to resolve $\mu_{\rm B}$ and $\mu_{\rm As}$ are the following.
First, their sum is the total or free energy per formula unit of perfect BAs determined from \textcolor{black}{\emph{ab initio} calculations,}
\begin{equation}
    \mu_{\rm A} + \mu_{\rm B} = E({\rm AB}),
    \label{sumeq}
\end{equation}
where $E$(AB) is the total energy per AB formula unit, which is also utilized in the HoF approach and is commensurate with the limits defined in Ref.~\onlinecite{zhang1991chemical}. 
However, unlike the HoF approach where either $\mu_{\rm B}$ and $\mu_{\rm As}$ is assumed equal to the elemental limit, the second constraint in our approach is that the overall stoichiometry is balanced by point defect concentrations, such that the ratio of the As and B number densities equals a predefined stoichiometry. 

\textcolor{black}{In a binary system with a given stoichiometry B$_{1+\delta}$A$_{1-\delta}$, the concentrations of all defects are practically calculated from the free energy of formation~,\cite{Mayer95,mishra2012native}
\begin{equation}
    C_X(T)=C_{\mathrm{s}}\theta_X \mathrm{exp}\left[-{F_X^f}/{\left(k_\mathrm{B}T\right)}\right],
    \label{eq:concentration}
\end{equation}
where $C_\mathrm{s}$ is the concentration of lattice sites in the perfect material, $\theta_X$ is the multiplicity of the configuration (which is different from 1 e.g. for dumbbell interstitials which can have different orientations), and $k_\mathrm{B}$ is the Boltzmann constant. 
These allow us to calculate the concentrations of As and B atoms in BAs by
\begin{eqnarray}
{{N_{\rm As}}\over{V}} &&={N\over V} -  C_{\rm V_{As}} - C_{\rm B_{As}} + C_{\rm As_B}+C_{{\rm As_i}}, 
\label{As_conc}
\nonumber \\
{{N_{\rm B}}\over{V}} &&={N\over V} -  C_{\rm V_B} - C_{\rm As_{B}} + C_{\rm B_{As}}+C_{ \rm B_i}, \label{B_conc}
\end{eqnarray}
where $V$ is the volume of the cell and $N$ is number of B or As atoms. 
For a given stoichiometry B$_{1+\delta}$As$_{1-\delta}$, the difference between the B and As concentrations needs to be $2\delta$, resulting in the equation 
\begin{equation}
C_{\rm V_{As}}+C_{\rm B_i}+C_{\rm B_{As}}-C_{\rm V_B}-C_{\rm As_i}-C_{\rm As_B}=2\delta{{N}\over{V}}.
\label{stoi_conc}
\end{equation}
For the total defect concentrations incorporating all charge states that is needed for Eq.~\ref{stoi_conc}, the concentrations in Eq.~\ref{B_conc} need to be calculated for all charge states and summed up into a total defect concentration. For larger energy differences between the charge states and at sufficiently low temperature, that procedure can be simplified by only calculating the concentration for the lowest-energy point defect which will dominate.} 

Replacing the concentrations in Eq.~\ref{stoi_conc} with the corresponding expressions from Eq.~\ref{eq:concentration} with the free energies of formation following the form of Eq.~\ref{eq:formation_energy} creates a second equation in addition to Eq.~\ref{sumeq}.
The combination of Eqs.~\ref{sumeq} and \ref{stoi_conc} allows solving for the chemical potentials. 
\textcolor{black}{For systems with more than two components, Eq.~\ref{stoi_conc} becomes a system of equations for pairwise stoichiometries. For the example of a ternary compound A$_m$B$_n$C$_l$, one ends up with two equations for $\delta_{\rm AB}$ and $\delta_{\rm AC}$ (which already fix the BC ratio too), where the three chemical potentials can then be solved in combination with the counterpart of Eq.~\ref{sumeq}. The detailed equations for this example are given in Sec.~S.3. in the SM.}
Since the free energies depend on temperature and Eq.~\ref{eq:formation_energy} adds the \textcolor{black}{Fermi level} into the equation, the chemical potential $\mu(T,E_F,\delta)$ as a function of $T$ and $E_F$ for the stoichiometric case ($\delta=0$) or a given non-stoichiometry $\delta$ can be determined. 

\textcolor{black}{In the limit of zero temperature where the Boltzmann factor becomes numerically inaccessible, the equation system collapses to the requirement that the formation energy of the lowest-energy stoichiometry balancing defects are equal.\cite{foiles_daw_1987}}
Specifically, one pair of defects resulting in B-rich and As-rich sample respectively should have the same formation energy, for example $E_f({\rm B_{As}})=E_f({\rm V_B})$ or $E_f({\rm B_{As}})=E_f({\rm As_B})$, or $E_f({\rm B_{As}})=E_f({\rm As_i})$, to name a few possible cases.
The solution of the chemical potentials from these postulation must satisfy that all other defects are ``irrelavant'', meaning that they have higher formation energies than the two stoichiometry-balancing defects.
As we will show later, the 0~K chemical potential obtained in this way agrees with the value extrapolated from finite temperatures very well.
One can already see that, at both 0~K and finite temperatures, the chemical potentials are obtained from the lattice stoichiometry in a self-consistent way, and all point defects \textcolor{black}{should} have non-negative formation energies, and thus physical equilibrium concentrations in our approach.

\textcolor{black}{While Eq.~\ref{stoi_conc} can be solved in principle for any off-stoichiometry $\delta$, as we show for the example of BAs in Sec.~S.4 and Fig.~S4 in the SM, the formation energy ($E_f$) of one of the point defects quickly saturates and approaches zero with increasing off-stoichiometry $\delta$, indicating that the chemical potential does not change upon alternating $\delta$ anymore. This then means that this zero-formation-energy defect is the one linked to the off-stoichiometry.
This connection has been discussed before for point-defect calculations for intermetallic compounds in the 1980's and 1990's,\cite{Mayer95,foiles_daw_1987,Schott97,hagen1998point} where the term ``constitutional defect'' was used for the zero-formation-energy point defect responsible for the off-stoichiometry, and the zero-formation energy was defined through the grand thermodynamic potential.
Practically, this means that, for the example of a BAs crystal, one of the potential defects that can cause a B-rich sample, i.e.\ a B$_{\rm i}$ interstitial, B$_{\rm As}$ antisite, or V$_{\rm As}$ vacancy, will have zero formation energy, while all other defects will have higher (positive) formation energies. If for an assumed constitutional defect, another defect is found to have negative formation energy, that defect needs to be the constitutional defect instead.
For the example of postulating that the B$_{\rm As}$ is the ``constitutional defect'' for B-rich BAs, the chemical potentials can be calculated by combining Eq.~\ref{sumeq} with the requirement that $E_f({\rm B_{As}})=0$ and substituting for $E_f({\rm B_{As}})$ the equivalent expression from Eq.~\ref{eq:formation_energy},
$E({\rm B_{As}})-E_{\rm perf}-\mu_{\rm B}+\mu_{\rm As}+Q(E_F+E_v).$}

\textcolor{black}{Beyond the constitutional limit, our canonical approach is identical to the grand canonical approach, except that the chemical potential does not necessarily have to be the same with the elemental chemical potential. 
In fact, it should not be equal because the chemical potential in a compound is always lower than that in the elemental limit, otherwise the compound will never form. Still, comparing the constitutional chemical potentials with the elemental limits gives additional information about the stability of stoichiometric and off-stoichiometric  compounds, as we discuss in the SM in Sec.~S.3 and Fig.~S3.}

While the condition of charge neutrality is discussed in the review of the HoF approach from 2014\cite{Freysoldt2014first} but typically not used in the reviewed body of work, it is a central point in the present work due to the Fermi-level dependence of our chemical potentials. 
While discussing the significance of charge neutrality in both approaches in detail in SM Sec.~S.1, we start here from the typical condition that the CNP is determined in an undoped crystal by adding up the charges on the defects weighted by their concentration $C_X$, calculated by Eq.~\ref{eq:concentration}, for each possible \textcolor{black}{Fermi level ($E_F$)} in the band gap, and equating them to the defect-induced carrier population,
\begin{equation}
    \Delta n_D(E) = n_D(E)-p_D(E) =\sum_X Q_X C_X
    \label{chex}
\end{equation}
where charges $Q_X$ on donor and acceptor point defects $X$ have opposite signs, such as $Q_{{\rm V}_\mathrm{B}^{+1}}= +1$ and $Q_{{\rm V}_\mathrm{As}^{-2}}= -2$. The resulting defect charge concentration, which we label $\Delta n_D(E)$, describes the electron or hole excess injected into the crystal from unbalanced charges on the defects. For extrinsic semiconductors, the extrinsic carrier concentration would have to be added into the equation, and electron vs.\ hole population would be subject to the mass action law. 

The charge neutrality point in BAs is thus defined by the condition that the excess charge carrier density $n(E)-p(E)=0$ within the condition of stoichiometry balancing at a given temperature. 
For Fermi levels below the CNP, the equilibrium point defects add extra hole and above the CNP, extra electron concentration, which we express by $\Delta n_D$ (positive for excess electrons, negative for holes). Fermi level and defect-induced carrier concentration can then be determined by standard semiconductor physics by combining Eq.~\ref{chex} with 
\begin{equation}
    E_{F_D} = E_{F_i}+k_B T\ln\left({{\Delta n_D + n_i}\over{n_i}}\right)
    \label{EFn}
\end{equation}
and solved self-consistently. 
\subsection{\label{sec:computational}Computational details}
The calculations were performed within density-functional theory (DFT) using the plane wave projector augmented-wave (PAW) method~\cite{blochl1994projector} as implemented in the VASP code~\cite{kresse1993ab,kresse1994ab}.
A cutoff energy of 560~eV and a BZ sampling grid of 6$\times$6$\times$6 Monkhorst-Pack k-point~\cite{monkhorst1976special} in a 64-atom super cell were used in this work. 
Monopole correction $e^{2}Q^{2}\alpha/(a_{0}\varepsilon$) ($\alpha$ is the Madelung constant, \textit{a$_0$} the lattice constant of the cubic cell and $\varepsilon$ the reported dielectric constant \cite{touat2006dynamical}) is applied to calculate energies of charged systems~\cite{windl1999first,Windl2004diffusion,Windl1998first}. 
While structural relaxation was performed with the Perdew-Burke-Ernzerhof (PBE) functional~\cite{perdew1996generalized}, the Heyd-Scuseria-Ernzerhof (HSE) hybrid functional~\cite{heyd2003hybrid} was employed for bandstructure calculations to get the correct band gap and for the final total-energy calculations to maintain maximum accuracy. 
This procedure (PBE relaxation plus HSE final energy) is computationally efficient but compares favorably to full HSE-relaxation, as we tested for the example of B$_\mathrm{As}$ antisite, the constitutional defect for B-rich BAs, where the calculated ``creation energy''~\protect\cite{Daw01}, i.e. the difference between the defect cell and the perfect supercell (without compensating chemical potentials) results in 20.60~eV with our procedure vs.~20.55~eV for full HSE relaxation, a difference of 0.24\%. 
The fraction of exact exchange in the HSE calculations was chosen as 0.16 to obtain an indirect zero-temperature band gap of 1.60~eV, since typical semiconductors have band gaps that shrink between 0~K and room temperature by approximately 0.1~eV~\cite{varshni1967temperature} and the measured room temperature gap for BAs is $\sim$1.50~eV~\cite{chu1974preparation}.
The quasiharmonic calculations of the vibrational free energy contributions were performed as outlined in Ref.~\onlinecite{luo2009first} as summarized in the SM, Section S.2. 

\section{\label{Sec:results}Results and Discussion}

\subsection{Intrinsic Fermi level and carrier concentration}
\label{Sec:SI_bandstructure}
\begin{figure}
\includegraphics[width=0.8\linewidth]{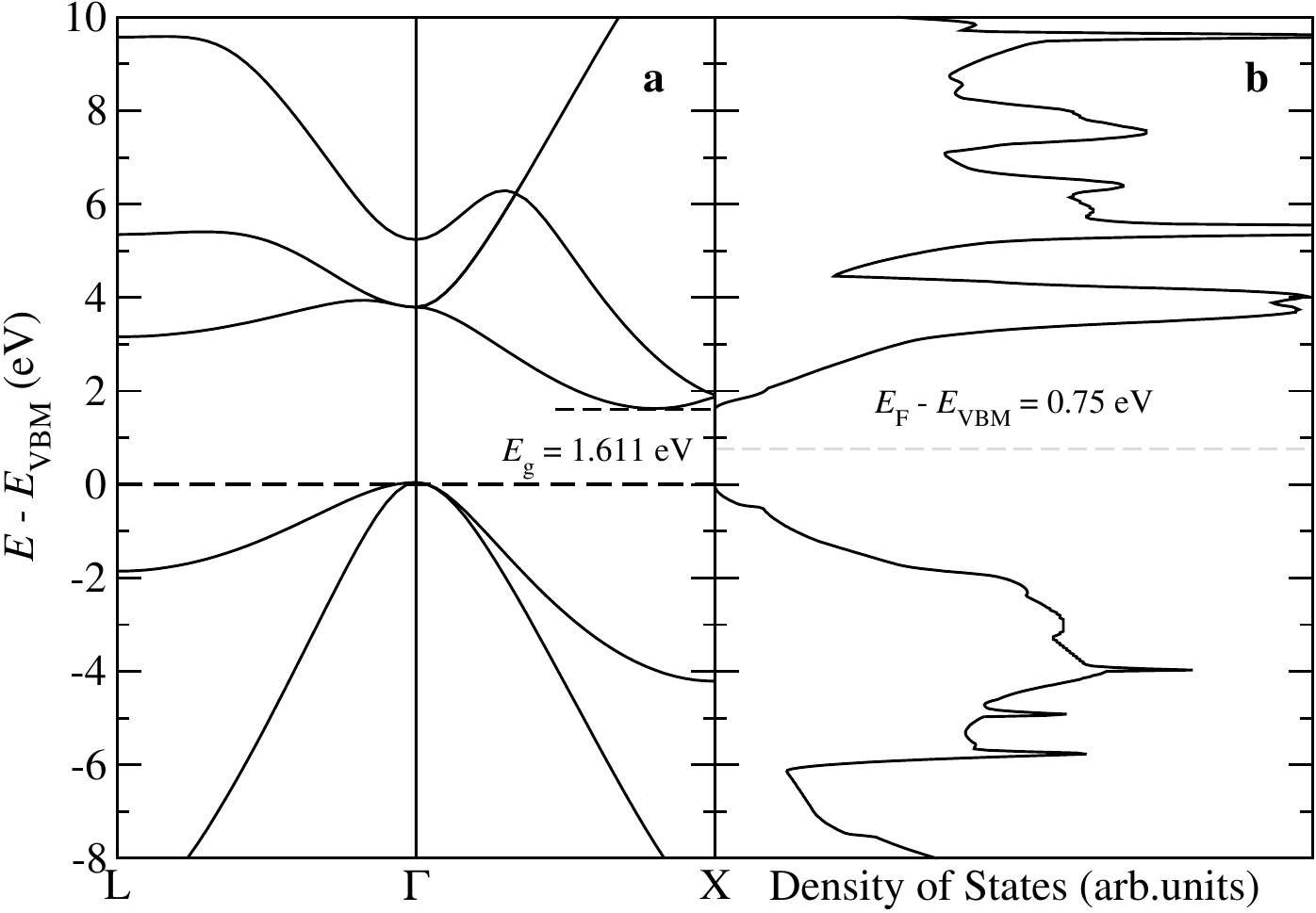}
\caption{Electronic structure of BAs calculated within DFT-HSE. \textbf{a} Band structure of the perfect cell with band gap labeled. \textbf{b} Density of states of perfect BAs. The Fermi level is calculated for a temperature of 1100~K using Eq.~(\ref{EFiEq}).}

\label{BSFig} 
\end{figure}
Figure~\ref{BSFig} shows band structure and density of states of perfect BAs.The effective mass of holes $m^*_p$ is obtained from an average of heavy hole and light hole and found to be 1.02~$m_0$\textcolor{black}{. Similar to that in silicon, the effective mass of the anisotropic conduction band minima is characterized by a longitudinal mass ($m_l=1.11m_0$) along the corresponding equivalent (100) direction and two transverse masses ($m_t=0.513m_0$) in the plane perpendicular to the longitudinal direction.\cite{van2004principles}} With the sixfold degenerate conduction band minimum,\cite{van2004principles}
\begin{equation}
    m^*_{n,{\rm dos}}=6^{2/3}\times(m_lm_tm_t)^{1/3}=2.19m_0.
\end{equation}
The intrinsic Fermi level ({0.75}~eV above the valence band edge) indicated in Fig.~\ref{BSFig}b has been calculated for a temperature of 1100~K, which is the temperature at which the BAs crystals of Lv \textit{et al.\ }have been synthesized~\cite{lv2015experimental}, using 
\begin{eqnarray}
E_\mathrm{i}&=&\frac{E_\mathrm{gap}}{2}+\frac{3}{4}k_\mathrm{B}T\mathrm{ln}(m_\mathrm{p}^*/m_\mathrm{n}^*)\nonumber \\
&=&\left[0.805-4.94\times 10^{-5} (T/{\rm K})\right]{\rm eV},
\label{EFiEq}
\end{eqnarray}
where $E_\mathrm{gap}$ denotes the band gap between conduction band minimum and valence band maximum , $k_\mathrm{B}$ the Boltzmann constant and $T$ the temperature.  
With the same ingredients, the intrinsic carrier concentration can be determined, 
\begin{equation}
 n_i(T) = \sqrt{N_cN_v}\exp\left(-\frac{E_g}{2k_B T}\right).
     \label{IntCarCon}
\end{equation}
With $E_g=1.611$~eV, $m^*_n=2.19m_0$, $m^*_p=1.02m_0$, and

\begin{eqnarray}
  N_c(T) &&= 2\left({{2\pi m_n^*k_BT}\over{h^2}}\right)^{3/2} =1.54\times10^{16} {\rm cm}^{-3} \left({{T}\over{\rm K}}\right)^{3/2}, \nonumber \\
 N_v(T) &&= 2\left({{2\pi m_p^*k_BT}\over{h^2}}\right)^{3/2} =4.97\times10^{15} {\rm cm}^{-3} \left({{T}\over{\rm K}}\right)^{3/2}, \nonumber
     \label{IntCarConNvNc}
\end{eqnarray}
we get
\begin{equation}
 n_i(T) = 8.85\times10^{15} \left(\frac{T}{\rm K} \right)^{3/2}\exp \left( -\frac{9351}{T/{\rm K}}\right) {\rm cm}^{-3}. \nonumber
     \label{IntCarConNum}
\end{equation}
$n_i$ has values of $6.56\times 10^{16}$~cm$^{-3}$ at 1100~K and $1.33\times 10^{6}$~cm$^{-3}$ at 300~K.

\subsection{Structure and energetics of point defects}

\begin{figure}
\includegraphics[width=0.9\linewidth]{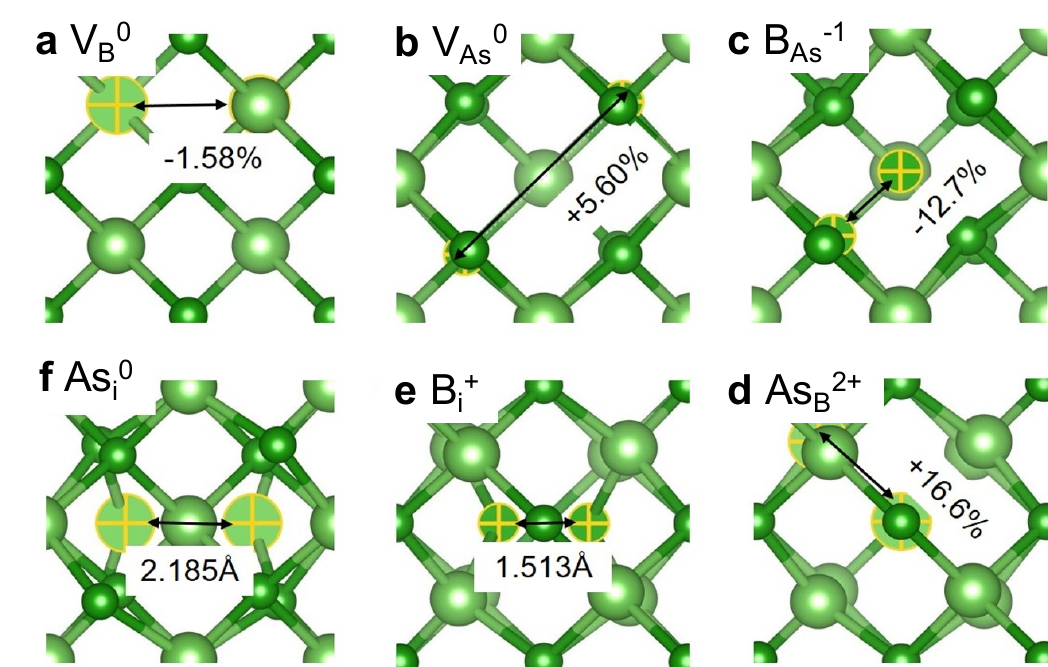}
\caption{Relaxed point defect structures for \textbf{a} B and \textbf{b} As vacancies, \textbf{c} B and \textbf{d} As antisites, \textbf{e} B and \textbf{f} As (100) dumbbell interstitials, viewed from the (100) direction. As (B) atoms are represented by larger (smaller) balls. }
\label{fig:structure} 
\end{figure}

Figure~\ref{fig:structure} shows the most stable structures of the various defects in their minimum-energy charge state at mid gap as labeled. The perfect cell is relaxed to have a lattice constant of {4.821~\AA} compared to the experimental value of {4.7802~\AA} reported in Ref.~\citenum{lv2015experimental}, with a B-As bond length of 2.088~\AA. 
The arrows in Figs.~\ref{fig:structure}a-d indicate changes in bond length as compared to perfect BAs, where we find that the lattice distortion from antisites is significantly larger than from vacancies. 
While B vacancies and antisites contract the lattice, their As counterparts expand it. 
For interstitials, the (100) dumbbell with identical species was found to be the most stable case for both boron and arsenic as compared to tetrahedral, hexagonal and (110) dumbbell interstitials. 
In Figs.~\ref{fig:structure}e and f, the arrows represent the bond length between the two dumbbell atoms. 
The relaxations were performed without enforcing crystallographic symmetry to allow for symmetry breaking (e.g. Jahn-Teller distortions or off-center dumbbells), which however was found to be very small.

\begin{figure}
\includegraphics[width=\linewidth]{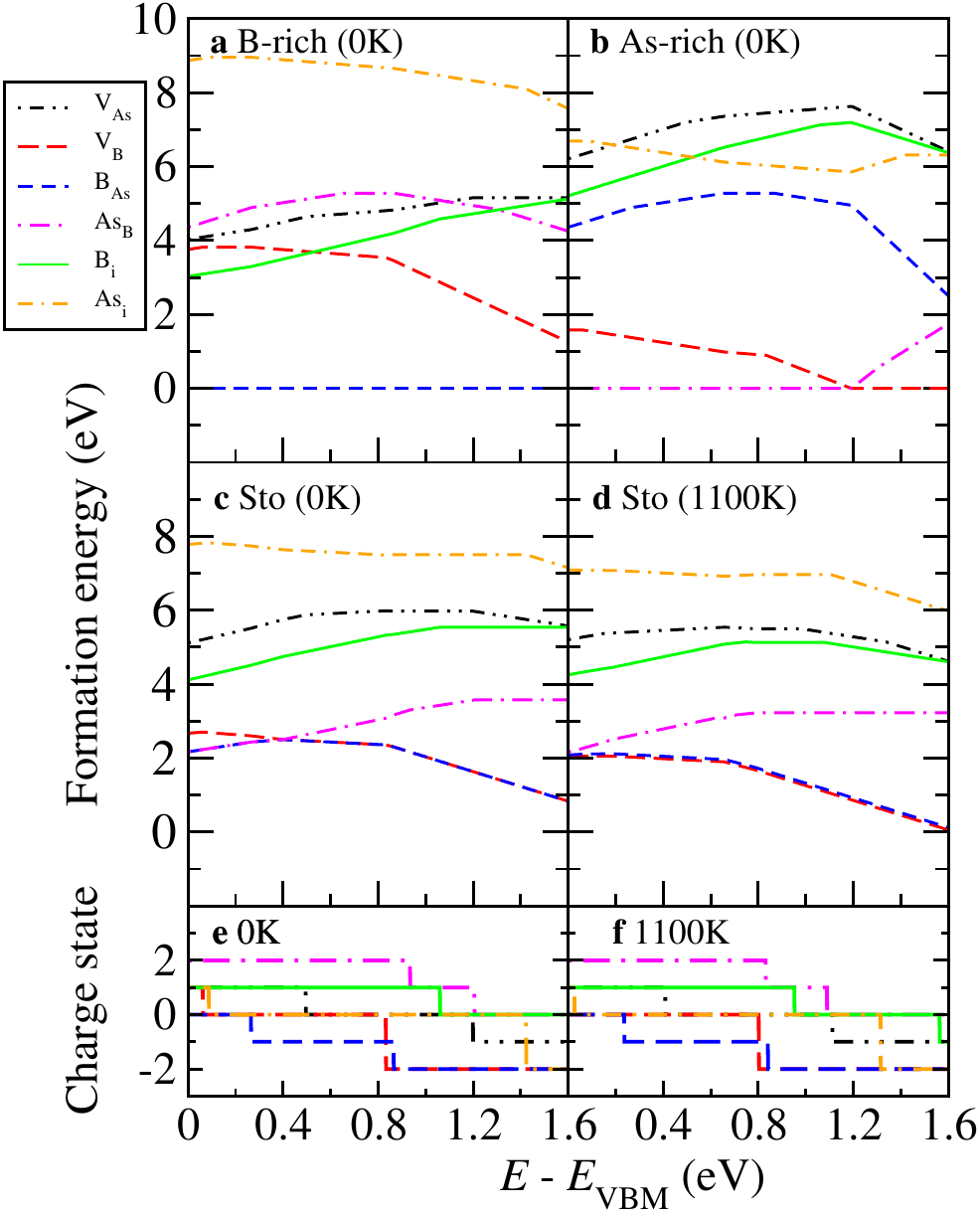}
\caption{Formation energies of native point defects as labeled for \textbf{a} B-rich, \textbf{b} As-rich, and \textbf{c} stoichiometric BAs at 0 K as a function of Fermi level. \textbf{d} Free energies of formation at the growth temperature (1100~K) of Ref.~\cite{lv2015experimental} for stoichiometric BAs as a function of Fermi level. {Minimum-energy charge states for the different defects as a function of Fermi level} {at} { {\bf e} 0 K and {\bf f} 1100 K for stoichiometric BAs.}  }
\label{fig:formation_energy} 
\end{figure}

The formation energies of all defects as a function of Fermi level relative to the top of the valence band $E_{\rm VBM}$ in both stoichiometric and off-stoichiometric cases are shown in Fig.~\ref{fig:formation_energy}a-d. 
We find B-antisites to be the constitutional defects in the B-rich case for the entire Fermi-level range (Fig.~\ref{fig:formation_energy}a), while for the  As-rich case, the As-antiste dominates up to $\sim 1.2$~eV, while for high Fermi levels, V$_{\rm B}$ takes over (Fig.~\ref{fig:formation_energy}b). 
The stoichiometry-balancing defects are independent of temperature except for the close vicinity of the valence band edge at low temperatures.  
At 1100~K, which is about the annealing temperature in Ref.~\onlinecite{lv2015experimental},  B$_{\rm As}$ and V$_{\rm B}$ balance each other to maintain stoichiometry over the entire Fermi-level range (Fig.~\ref{fig:formation_energy}d), while the zero-temperature calculations predict stoichiometry balancing between As$_{\rm B}$ and B$_{\rm As}$ instead of V$_{\rm B}$ in the severe $p$-type region (Fig.~\ref{fig:formation_energy}c).

\subsection{Charge neutrality}
\label{Sec:main_CNP}

In order to explore the relation between charge neutrality, Fermi level and non-stoichiometry in our results, we start from a short discussion of the  charge neutrality point (CNP). 
The term CNP is historically used to indicate the ``pinning'' of the Fermi level at interfaces of a semiconductor with other materials~\protect\cite{Spicer79} and determines the change of the Fermi level with increasing defect concentrations.\protect\cite{Brudnyi95} 
Thus, it is the essential link between point defect concentrations and measurable electronic properties and is explained in more detail in SI Sec.~S.1.

Within the HoF approach, where the chemical potentials are Fermi-level independent, the charge of a defect can be determined by its slope in the plot of formation energy vs.\ Fermi level. This allows to identify the zero-temperature CNP from the lowest-energy intersection between two formation energy curves with slopes of the same magnitude but opposite signs. This is important, since the formation energies at the CNP are the in principle only sensible ones as opposed to other Fermi level positions, where the formation energies can even be negative (see Sec.~S.1).
This method would not work within our framework, where the chemical potentials are Fermi-level dependent, while the favorable charge state of a certain defect is the one leading to the lowest formation energy. Thus, the slope of the formation energy curves is not related to the defect charge. For example, the B$_{\rm As}$ antisite is neutral at $E-E_{\rm VBM}<0.265$~eV (Fig.~\ref{fig:formation_energy}e), but the formation energy does not have a zero slope at the same range (Fig.~\ref{fig:formation_energy}c). Therefore, the charges need to be separately recorded and reported.
Figures~\ref{fig:formation_energy}e-f show the charge states for the native point defects at both 0~K and 1100~K, where the ionization levels $\varepsilon(q_1/q_2)$ for the different defects can be identified from the steps in the charge vs.\ Fermi level plots.

\begin{figure}
\includegraphics[width=0.9\linewidth]{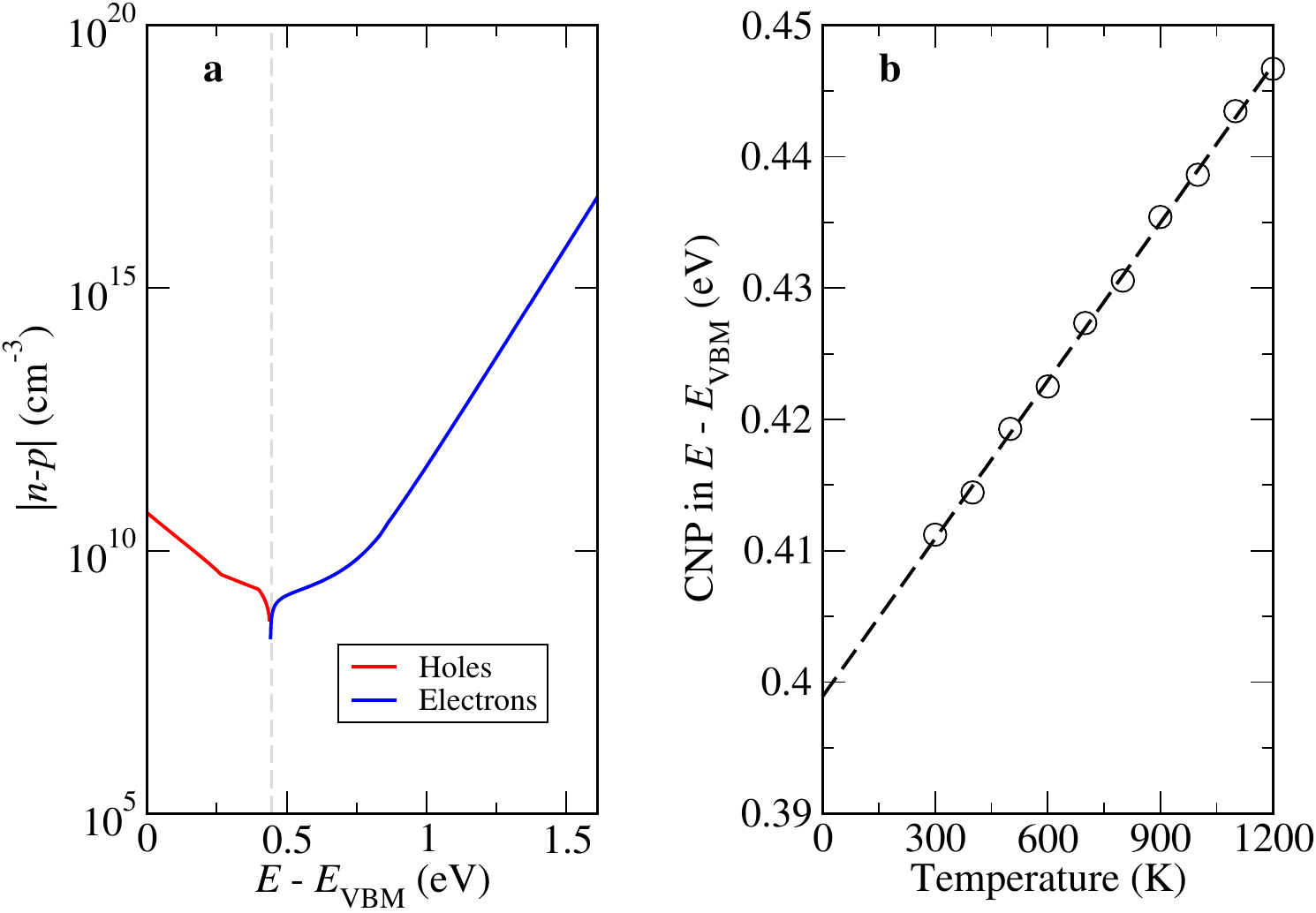}
\caption{ \textbf{a}. Charge neutrality point determined from the zero value of charge excess, due to charged point defects for stoichiometric BAs at {1100~K}. 
The CNP is drawn as grey dashed line, hole and electron excess are drawn in red and blue lines, respectively. 
\textbf{b}  Charge neutrality points as a function of temperature (circles) and linear fit (dashed line) to determine the charge neutrality point at zero temperature by extrapolation. 
All chemical potentials were determined by the condition of stoichiometry conservation.}
\label{ChgCarFig}
\end{figure}

In the spirit of stoichiometry balancing, the CNP is determined quantitatively by Eq.~\ref{chex} using the defect concentrations as defined in Eq.~\ref{eq:concentration}. 
Figure~\ref{ChgCarFig}a shows the electrons or holes injected from all possible native point defects in stoichiometric BAs at 1100~K, unbalanced by the defect charges. 
The excess charge density $|n-p|$ will diverge at the CNP, indicated by the grey line in Fig.~\ref{ChgCarFig}a.
For stoichiometric BAs, as calculated at 1100~K in Sec.~\ref{Sec:SI_bandstructure}, the intrinsic Fermi level (Eq.~\ref{EFiEq}) is 0.75~eV and the intrinsic carrier concentration (Eq.~\ref{IntCarConNum}) is $6.56\times 10^{16}$~cm$^{-3}$, while the excess charge concentration is $|n-p|\sim 6.9\times 10^{9}{\rm cm}^{-3}$, seven orders of magnitude lower.
With Eq.~\ref{EFn}, the Fermi level stays almost in the middle of the gap as it should, since the overall Fermi energy in a bulk semiconductor is a thermodynamic average and the excess carriers from the charged defects could barely shift it significantly. This situation can be very different in non-stoichiometric BAs, where the concentration of the constitutional defects can be in the percent range, potentially resulting in a significant shift of the Fermi level. The connection between stoichiometry, CNP, and resulting \textcolor{black}{Fermi level} is thus a straightforward outcome of our approach, while in traditional HoF work it is usually not possible due to the undefined relationship between stoichiometry and chemical potentials.
In the following, we will use the CNP in the traditional sense, where it refers to the charge-balancing Fermi level \emph{only} from the native point defects.\protect\cite{Spicer79}

Next, we compare the CNP from stoichiometry balancing to that determined from HoF plots. Due to the instability of defect concentrations at extremely low temperatures, we extrapolate the CNP from a series of finite-temperature results. With that, we find it to be $0.40$~eV above the valence band edge at 0~K (Fig.~\ref{ChgCarFig}b).
Previous work at 0~K has also found it to be in the $p$-type regime below mid gap at $\sim0.6$~eV above the valence band edge.\cite{zheng2018antisite}
Possible reasons for the slight discrepancy include that Ref.~\onlinecite{zheng2018antisite} assumes $\mu_{\rm As}$ to be at its elemental limit, the inclusion of zero-point energies in our results which influence the defect concentration balancing, and the different HSE mixing parameters in the HSE functional, 0.25 in Ref.~\onlinecite{zheng2018antisite} vs.~0.16 here (see Sec.~\ref{sec:computational}).

\textcolor{black}{Further,} we compare the \textcolor{black}{intrinsic-defect-doping} in the different stoichiometry limits with the intrinsic carrier concentration \textcolor{black}{ and extrinsic-impurity-doping.} 
\textcolor{black}{The atomic concentration of B/As in a stoichiometric sample is $\sim3.67\times10^{22}$~cm$^{-3}$, where Eq.~\ref{IntCarConNum} calculates the intrinsic carrier concentration to be $6.56\times10^{16}{\rm cm}^{-3}$ at 1100~K and the thermal point defects are predicted to produce $n$-type doping} with an electron concentration of  $6.9\times10^{9}$~cm$^{-3}$ at 1100~K, seven orders of magnitude smaller than the intrinsic carrier concentration, which then leaves the Fermi level at almost mid gap as expected for a stoichiometric semiconductor.
\textcolor{black}{However, for the heavily B-rich sample from Ref.~\citenum{lv2015experimental}, assuming all off-stoichiometry is caused by the B$_{\rm As}^{-1}$ antisite as constitutional defect, the 2.8\% As-deficiency corresponds to a B$_{\rm As}^{-1}$ concentration of approximately $5.14\times10^{20}$~cm$^{-3}$  acting as acceptors.}
\textcolor{black}{In the non-degenerate region, the carrier concentration can be estimated by $N_A/\{1+4{\rm exp}[(E_A-E_v)/k_BT]\}$ for holes, where $N_A=5.14\times10^{20}$~cm$^{-3}$ is now the acceptor concentration and $E_A=0.265$~eV is the ionization energy (Fig.~\ref{fig:formation_energy}e). 
Thus the hole concentration from self-doping is $\sim 4.45\times10^{15}$~cm$^{-3}$ at 300~K, and $\sim 7.68\times10^{18}$~cm$^{-3}$ at 1100~K.}
In combination with the low enthalpy of formation of  B$_{12}$As$_2$, which puts the BAs phase above the convex hull and makes it much less likely to grow an As-rich sample (SM Sec.~S.5), this leads to the conclusion that standard growth techniques most likely produce \textit{p}-type boron rich samples. 
\textcolor{black}{Although the hole concentration from the B$_{\rm As}^{-1}$ antisites alone are not as high as the magnitude of $10^{19}$~cm$^{-3}$ from earlier works which suggests that Si/C impurities are indeed dominant in the $p$-type conduction,~\cite{chae2018point,lyons2018impurity,meng2020pressure}
our results indicate that even without impurities, the as-grown samples would still be $p$-type, although to a lesser degree.
Our canonical native point defect formation energies thus enable a quantitative analysis on the origin of holes from the native vs.\ extrinsic defects, thus allowing the assessment of the effect of sample purity vs.\ sample stoichiometry. 
To this end, our canonical approach opens up new opportunities to quantitatively assess the self-doping in non-stoichiometric samples, especially important in materials where intrinsic point defects can have a smaller ionization energy.}

\subsection{Free energies and thermodynamic stability}
\label{Sec:main_murange}
\begin{figure}
\includegraphics[width=\linewidth]{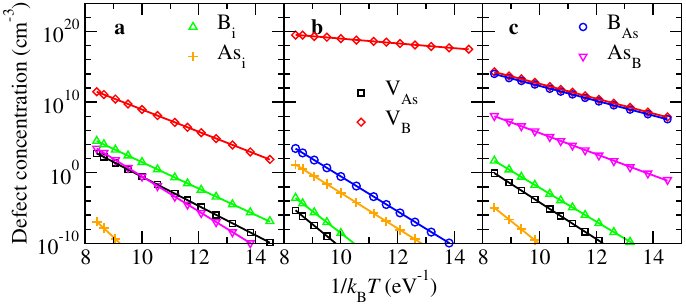}
\caption{ Thermal defect concentrations vs. 1/$k_\mathrm{B}T$ between 800 and 1380 K in steps of 20 K in \textbf{a} B-rich, \textbf{b} As-rich and \textbf{c} stoichiometric BAs, determined from free energies calculated within the quasi-harmonic approximation. 
Every second data point plotted for clarity. 
Solid lines are linear Arrhenius fits. In cases \textbf{a} and \textbf{b}, the constitutional defects are not shown.  }
\label{fig:free_energy}
\end{figure}

Since we have calculated the full free energies of formation, we now plot the various defect concentrations as a function of (inverse) temperature in an Arrhenius plot as shown in Fig.~\ref{fig:free_energy} for intrinsic BAs at mid gap \textcolor{black}{in order to examine if zero-temperature formation energies are a viable way to estimate defect concentrations and thus can reasonably well substitute for the formation enthalpies $H_f$ from the full free energies of formation (independently of how the chemical potentials are determined).} 
We choose to examine them at mid gap because as discussed earlier, the intrinsic carriers  dominate in the chosen temperature range and leave the overall Fermi level close to mid gap, with only small shifts from the charged defects as long as we assume small non-stoichiometries.
All lines are in very good approximation linear, and with \textit{F$_f$ = H$_f$ $-$ TS$_f$}, we can use a linear fit to determine formation enthalpy and entropy for each defect, shown in Table~\ref{table:enthalpy}. 
These formation enthalpies contain contributions from zero-point energies and compensate for non-linearities from the vibrational entropy contribution, and thus cannot \textcolor{black}{a priori} be expected to be identical to the formation energies calculated from zero-temperature total energies.
The number of digits shown in the enthalpy column indicates the degree of linearity of the fit, and the error bars shown limit the 95\% confidence interval. The concentrations follow an Arrhenius temperature dependence to an excellent degree, although that does not necessarily have to be the case for the type of calculations performed here as previously found~\cite{stoddard2005ab,grabowski2009ab,Grabowski2010formation,glensk2014breakdown}. 
We approximate the configurational entropy per atom in the dilute limit by~\cite{Freysoldt2014first}
\begin{equation}
    S_f^{\rm conf}=k_{\rm B}[c-c{\rm ln}(c)+c{\rm ln}(\theta_X)],
\end{equation}
where $c=n/N$ with $n$ being the number of point defects of a specific type and $N$ is the number of lattice sites; $\theta_X$ is again the configurational degree of freedom. 
For the most relevant defects, i.e. antisites and vacancies, $\theta_X = 1$.
Even if we use the simulated supercell to account for defect concentration (1 defect in 32 lattice sites thus $c\sim 3\%$), which is usually exaggerated, the configurational entropy is about $S_f^{\rm conf}=0.1396 k_{\rm B}$, corresponding to $-0.012$~eV at 1000~K.
This contribution is negligible compared to the vibrational entropy listed in Table~\ref{table:enthalpy}.

While the agreement between calculated formation energies and fitted formation enthalpies is in general good with an average deviation of 5\%, the considerable variations in the \textcolor{black}{(vibrational)} formation entropies shown in Table~\ref{table:enthalpy} cause some larger deviations up to 14\%. 
While the sequence and relative importance of nearly all point defects is not dependent on temperature, there is one exception: the As$_{\rm B}$ antisite dominates close to zero temperature in proximity to the valence band edge over V$_{\rm B}$, while V$_{\rm B}$ dominates otherwise in most of the relevant temperature and Fermi level range as the stoichiometry balancing defect. 
In combination with the significant variations in formation entropy in Table~\ref{table:enthalpy} and thus different temperature dependence, our results suggest that a calculation of free energies is indeed an important step to determine the dominant defects at finite and especially high annealing temperatures, while the calculation of zero-temperature formation energies give a reasonable estimate for the general trends and for identifying the most relevant defect types.
\textcolor{black}{While this was done for the specific choice of chemical potentials proposed in our paper (stoichiometric chemical potential from the canonical assumption for $N_{\rm B} = N_{\rm As}$; B- and As-rich cases from the assumption that the respective constitutional defects have zero formation energies), this comparison could have been done the same way for HoF-determined chemical potentials. }

The combination of Figs.~\ref{fig:formation_energy}a and \ref{fig:free_energy}a shows that for B-rich material, the highest concentration of thermal, non-constitutional defects is below 10$^8$~cm$^{-3}$, and thus all thermal defects should be negligible when calculating materials properties. 
This means the only relevant defect in B-rich material, which is the most relevant case for comparison to experiment as we just showed, is the boron antisite. 
The situation is different for As-rich BAs (Figs.~\ref{fig:formation_energy}b and \ref{fig:free_energy}b), where boron vacancies can reach concentrations in the 10$^{20}$~cm$^{-3}$ range, which may require including them into transport and carrier-concentration calculations in addition to the As$_{\rm B}$ antisites, while other defects still have negligible concentrations. 
For stoichiometric BAs {(Figs.~\ref{fig:formation_energy}c-d and ~\ref{fig:free_energy}c)}, we find that V$\mathrm{_B}$ and B$\mathrm{_{As}}$ balance the stoichiometry, with the equilibrium concentration of B vacancies twice that of B$_{\rm As}$ antisites, since one B$_{\rm As}$ needs two B vacancies to result in perfect stoichiometry. 
V$\mathrm{_B}$ and B$\mathrm{_{As}}$ concentrations have relatively low concentrations of $\sim$10$^{10}$~cm$^{-3}$ at elevated temperatures and significantly less around room temperature, giving them negligible relevance for electronic properties and thermal conduction in stoichiometric BAs and in agreement with the recent findings of close-to-ideal thermal conductivity in apparently stoichiometric BAs.\cite{kang2018experimental}

\begin{table}
\caption{Summary of zero-temperature formation energies $E_f$ and Arrhenius-fitted formation enthalpies $H_f$ at mid gap for native point defects in cubic BAs, along with their relative differences $\Delta E_\mathrm{rel}$ and Arrhenius-fitted formation entropies $S_f$.}

\begin{ruledtabular}
\label{table:enthalpy}
\setlength\extrarowheight{5pt}
\begin{tabular}{cccccc}
 Defects& & $E_f\mathrm{(eV)}$&$H_f\mathrm{(eV)}$
 &$\Delta E_\mathrm{rel}$(\%)&$S_f$($k_\mathrm{B}$)\\
\hline
&$\mathrm{V_{As}^0}$& 4.79 & 4.773$\pm$0.001 & 0.42 &1.39 \\
&$\mathrm{V_{B}^0}$& 3.36& 3.628$\pm$0.003 & 8.04 &11.8 \\
B-rich&$\mathrm{B_{As}^{-1}}$& 0& 0 & 0 & $-$\\
&$\mathrm{As_{B}^{2+}}$& 5.36& 5.702$\pm$0.004 & 6.74 &10.8 \\
&$\mathrm{B_{i}^{+}}$& 4.20& 4.312$\pm$0.01 & 2.62 &0.426 \\
&$\mathrm{As_{i}^0}$& 8.45& 8.635$\pm$0.001 & 2.25 &10.5 \\ \hline
&$\mathrm{V_{As}^0}$& 7.47 & 7.624$\pm$0.001 & 2.01 &6.79 \\
&$\mathrm{V_{B}^0}$& 0.68& 0.7763$\pm$0.0014 & 14.1 &6.46 \\
As-rich&$\mathrm{B_{As}^{-1}}$& 5.35& 5.702$\pm$0.004 & 6.54 & 10.08\\
&$\mathrm{As_{B}^{2+}}$& 0& 0 & 0 & $-$\\
&$\mathrm{B_{i}^{+}}$& 6.88& 7.163$\pm$0.011 & 4.07 &5.83 \\
&$\mathrm{As_{i}^0}$&5.77& 5.784$\pm$0.001 & 0.17 &5.14 \\ \hline
&$\mathrm{V_{As}^0}$& 5.98 & 5.982$\pm$0.001 & 0 &5.12 \\
&$\mathrm{V_{B}^0}$& 2.37& 2.418$\pm$0.003 & 2.04 &8.15 \\
Stoichi-&$\mathrm{B_{As}^{-1}}$& 2.37& 2.418$\pm$0.003 & 2.04 & 7.45\\
ometric&$\mathrm{As_{B}^{2+}}$& 3.04& 3.464$\pm$0.018 & 13.9 & 2.55\\
&$\mathrm{B_{i}^{+}}$& 5.29& 5.521$\pm$0.01 & 4.35 &4.15 \\
&$\mathrm{As_{i}^0}$&7.50& 7.442$\pm$0.001 & 0.8 &5.05 \\

\end{tabular}
\end{ruledtabular}
\end{table}
At this point, it is important to note that the chemical potentials numerically solved in a self-consistent manner as a function of stoichiometry, temperature and \textcolor{black}{Fermi level} $\mu(E_F,T,\delta)$ actually fulfill the stability criterion and do not exceed the boundaries set by the elemental limits (for more details see SM Sec.~S.3).
It is well known that increasing the temperature will facilitate defect formation, which in our temperature-dependent approach manifests itself in the fact that the temperature-dependent elemental limits for the chemical potentials form a wider range at elevated temperatures, where the numerical solution for the chemical potential stays well within the elemental limits for the entire \textcolor{black}{Fermi level} range (Fig.~S3 in SI). 
Importantly, the expansion of the chemical potential range at higher temperatures agrees with the phase stability analysis regarding the BAs and the B$_{12}$As$_6$ phases on a convex hull (Fig.~S5 in SI).
Therefore, this method with stoichiometry-linked chemical potentials can also be useful for tackling for example solubility ranges and phase stability problems in alloys or intermetallic compounds.

\section{\label{Sec:conclutions}Conclusions}
In summary, we have presented a new approach to calculate defect formation energies in non-metals based on stoichiometry-linked chemical potentials and shown results for the energetics of native point defects in cubic boron arsenide. 
We have demonstrated how, under the conditions of thermodynamic equilibrium and overall stoichiometry, chemical potentials are functions of temperature, \textcolor{black}{Fermi level}, and stoichiometry, and have demonstrated how they can be determined for our demonstration system BAs. With those, we have calculated the formation energies and equilibrium concentrations of the native point defects, which included vacancies, antisites and interstitials. 

We find that antisites are the constitutional defects in B-rich and to a large degree in As-rich material, while B$_\mathrm{As}$ antisites and B vacancies balance to keep stoichiometry over most conditions, while As$_{\rm B}$ replaces V$_{\rm B}$ in extremely $p$-type material at low temperatures. 
In addition, full free energies of formation were calculated for a range of temperatures, and we find that considerable differences up to 14\% exist between Arrhenius fitted formation energies and 0~K total-energy values. 

This work takes an important step further to practically include charge neutrality in a gapped material from the charged point defects in the analysis, which to a very large degree \textcolor{black}{has been left ambiguous} in most of the HoF literature. 
We have shown the critical significance of sensible charge neutrality analysis for our demonstration system, where  \textcolor{black}{we find that the as-grown B-rich samples are already intrinsically $p$-type, independent of the external doping through Si or C found to dominate doping in previous work.\cite{chae2018point,lyons2018impurity,meng2020pressure}}

\textcolor{black}{While the widely used HoF approach seems to be at the first glance transparent and straightforward, some care is required when interpreting the results.
If that care is taken, the HoF is a quick way to get an idea about dominant point defects and its general influence on the Fermi level, albeit not in a quantitative way.}
Our stoichiometry-balancing approach allows to correlate chemical potentials with stoichiometry in a physically rigorous way for a real material, and therefore allows a quantitative insight into point defect populations instead of the qualitative picture that is  gained from common HoF calculations, which in turn enables a quantitative connection to materials properties in semiconductors. 
Finally, we would like to point out that this method, when collapsed to a zero-gap system, directly connects to previous approaches for point-defect calculations in alloys and intermetallic compounds,\cite{Mayer95,foiles_daw_1987,Schott97,mishra2012native,hagen1998point} where it allows a quantitative evaluation of solubilities and point defects, especially the constitutional ones.

\section{Data Availability}
The data are available from the corresponding author on request.
\section{acknowledgments}
This work was funded by the Air Force Office of Scientific Research under award number FA9550-14-1-0322. Y.\ W.\ also acknowledges partial funding from the Center for Emergent
Materials, an NSF MRSEC (DMR 1420451). All high-performance computations were performed on machines of the Ohio Supercomputer Center, grant number PAS0072.
\section{Supplementary material}
See supplemental material for detailed discussions on charge neutrality point, quasiharmonic approximation, \textcolor{black}{chemical potential saturation}, as well as thermodynamic equilibrium and phase stability of BAs.

\providecommand{\noopsort}[1]{}\providecommand{\singleletter}[1]{#1}%

\end{bibunit}

\onecolumngrid
\begin{bibunit}
\clearpage

\begin{flushright}
 \Large{Supplemental Material}
\end{flushright}

\section*{Table of Contents}

\noindent
Section S.1 Charge Neutrality Point

Figure S1. Fermi-level dependent formation energies of point defects in Al$_2$O$_3$ and BAs from previous work.\

\noindent
Section S.2 Quasiharmonic Approximation

Figure S2. Convergence test for free energies in BAs.\

\noindent
Section S.3 Thermodynamic Equilibrium of BAs with Equilibrium Point Defects.\

Figure S3. Chemical potentials for As in stoichiometric BAs at 0~K and finite temperature.\

\noindent
Section S.4 Transition from canonical to grand canonical at off-stoichiometric limit

Figure S4. Stoichiometry (As/B atomic ratio) dependent defect formation energies in BAs.

\noindent
Section S.5 Phase Stability of Cubic Stoichiometric BAs

Figure S5. Convex hull plot of the formation energy of the phases in the B-As system.

\noindent

\section{Charge Neutrality Point From Heat-of-formation Method}
\label{Sec:SI_CNP}

\begin{figure}
\includegraphics[width=0.8\linewidth]{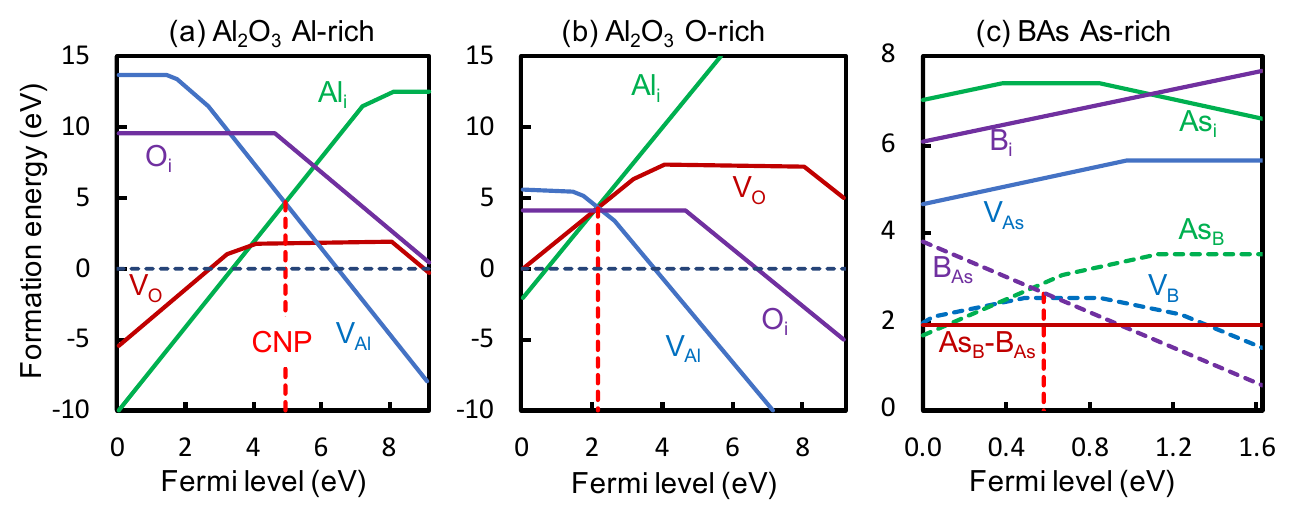}
\caption{ Fermi-level dependent formation energy of native point defects in (a) Al-rich and (b) O-rich Al$_2$O$_3$, calculated within the heat-of-formation approach, redrawn after the results from Ref.~\protect\onlinecite{Choi13}. (c) Fermi-level dependent formation energy of native point defects in BAs, calculated within the heat-of-formation approach for the As-rich case, redrawn after the results from Ref.~\protect\onlinecite{zheng2018antisite}.}
\label{CNP-Fig}
\end{figure}

Upon irradiation, the Fermi level of semiconductors starts to shift due to the point defects created, until it eventually stabilizes at a ``limiting'' value or neutrality level, from where it does not change with further irradiation \cite{Brudnyi95}. Also, the CNP can pin the Fermi level at interfaces, which can lead to unexpected band line-ups~\cite{Spicer79}.
Thus, the charge neutrality point (CNP) is a crucial ingredient for the interpretation of point defect energies in binary compounds. As an example for the meaning of the CNP in point-defect calculations,  Fig.~\ref{CNP-Fig}(a) shows the Fermi-level dependent formation energy of native point defects in Al$_2$O$_3$, redrawn after the results from Ref.~\onlinecite{Choi13}. The formation energies were calculated within the heat-of-formation approach for the Al-rich case, i.e.~the chemical potential of Al, $\mu_{\rm Al}$, was set equal to the atomic energy in elemental aluminum, and $\mu_{\rm O} = \left[ E_{\rm tot}({\rm Al}_2{\rm O}_3)-2\mu_{\rm Al}\right]/3$.

We have added into the figure a black dashed line at zero formation energy. As can be seen, a lot of the calculated formation energies are in part significantly below zero, which defies the definition of formation energy in a stable crystal and thus is unphysical. While it is usually not discussed explicitly, there is actually only one Fermi-level position in this graph where the results within this approach are sensible, which is at the Fermi level value that defines the CNP, indicated by the red dashed line in the figure. At that Fermi level value, the line for the lowest-energy point defect with positive slope intersects the lowest-energy negative-slope line with the same absolute value of slope. Since the slope in this fixed-chemical{-}potential approach corresponds to the point defect?s charge, at this point, positively and negatively charged point defects have the same formation energy and thus the same concentration, compensate each other?s charge and thus are used to define what Fermi level is sensible for the chosen chemical potential. Here, Al$_i$ (slope $+3$) and V$_{\rm Al}$ (slope $-3$) intersect at 4.9~eV, which thus defines the CNP. All the regions with the unphysical negative formation energies are thus irrelevant. At the CNP, the stable point defect (lowest energy) is V$_{\rm O}^0$ {with zero charge}. If thus V$_{\rm O}^0$ is the predominant point defect, the crystal will be Al-rich {without losing charge neutrality}, which is the premise of these results, and thus we end up finally with a piece of physical and valuable information and have identified the constitutional point defect and its charge. One could pick a different set of chemical potentials between the elemental limits and end up with a different CNP Fermi level. More specifically, in Fig.~\ref{CNP-Fig}(b) for the O-rich case, V$_{\rm O}$ (slope +2) intersects with V$_{\rm As}$ (slope -2) at around 2~eV, where the defect having the lowest formation energy is the uncharged oxygen interstitial, corresponding to an O-rich crystal.  In a similar fashion, as long as only single point defects are considered, Ref.~\onlinecite{zheng2018antisite} finds for the As-rich case of BAs, that at the charge neutrality point, which is 0.5-0.6 eV above the valence band edge, the neutral boron vacancy has the lowest energy and thus is constitutional defect, resulting in an As-rich crystal in agreement with the premise (Fig.~\ref{CNP-Fig}(c)). However, the ad hoc addition of the antisite pair to the complete set of single point defects now suddenly seems to suggest that the antisite pair would be the constitutional defect. Since however the antisite pair preserves the perfect stoichiometry, a contradiction is now created to the assumption of an As-rich crystal, suggesting a methodological problem by singling out one ad hoc cluster to throw into the mix. Independent of that, all three examples show that chemical potentials are clearly coupled to the Fermi level once a stoichiometry is chosen, also in the common heat-of-formation approach. While the CNP determined in this way can explain qualitatively the effect of Fermi-level pinning described above, it does not allow for a quantitative connection between stoichiometry and Fermi level, and does not connect to the well-known fact that a crystal with very few point defects has its Fermi level at mid gap.

%

\section{Quasiharmonic Approximation}
\label{Sec:SI_QHA}

As described in Ref.~\onlinecite{luo2009first}, we used the quasiharmonic approximation (QHA) with VASP input for energies and phonon frequencies to calculate the vibrational contribution to the free energy, which approximates the free energy by
\begin{equation}
    F(p,T)=H(p)+{1\over{4\pi}}\int g\left[\omega(p)\right] h\omega(p) d\omega +k_BT\int g\left[\omega(p)\right] \ln\left[1-\exp\left(-{{h\omega(p)}\over{2\pi k_BT}}\right)\right]d\omega 
\end{equation}
where $\omega$ denotes {phonon} frequencies, $g$ is the phonon density of states (PDOS), the second term on the right hand side is the zero-point energy, and the third term is the temperature-dependent contribution. In order to calculate the PDOS, we determine the normal modes in the defect supercell and equivalent perfect supercell using density-functional perturbation theory within VASP with the same settings as for the PBE calculations. The PDOS is calculated from a sum over the calculated $\Gamma$-point phonons, where each of them is broadened by a Gaussian of width 0.5 THz. As we have shown in Ref.~\onlinecite{luo2009first}, this rather fast method agrees typically within a few percent with a free-energy calculation from the PDOS calculated from a full integration by tetrahedron method over the Brillouin zone. This is expected since the $\Gamma$-point is a ``special'' k-point as defined first by Chadi and Cohen \cite{Chadi73} that is also frequently used to approximate electronic Brillouin-zone integration, and the accuracy of such single or multiple special points has been discussed in the childhood of DFT calculations extensively by e.g.\ Baldereschi \cite{baldereschi73} or Monkhorst and Pack \cite{monkhorst1976special}, respectively. Since all structures are relaxed to zero pressure, the enthalpy term $H(p)$ is equal to the total energy. As additional approximation for computational efficiency, the effect of thermal expansion on the phonon frequencies is neglected. 

\begin{figure}
\includegraphics[width=0.5\linewidth]{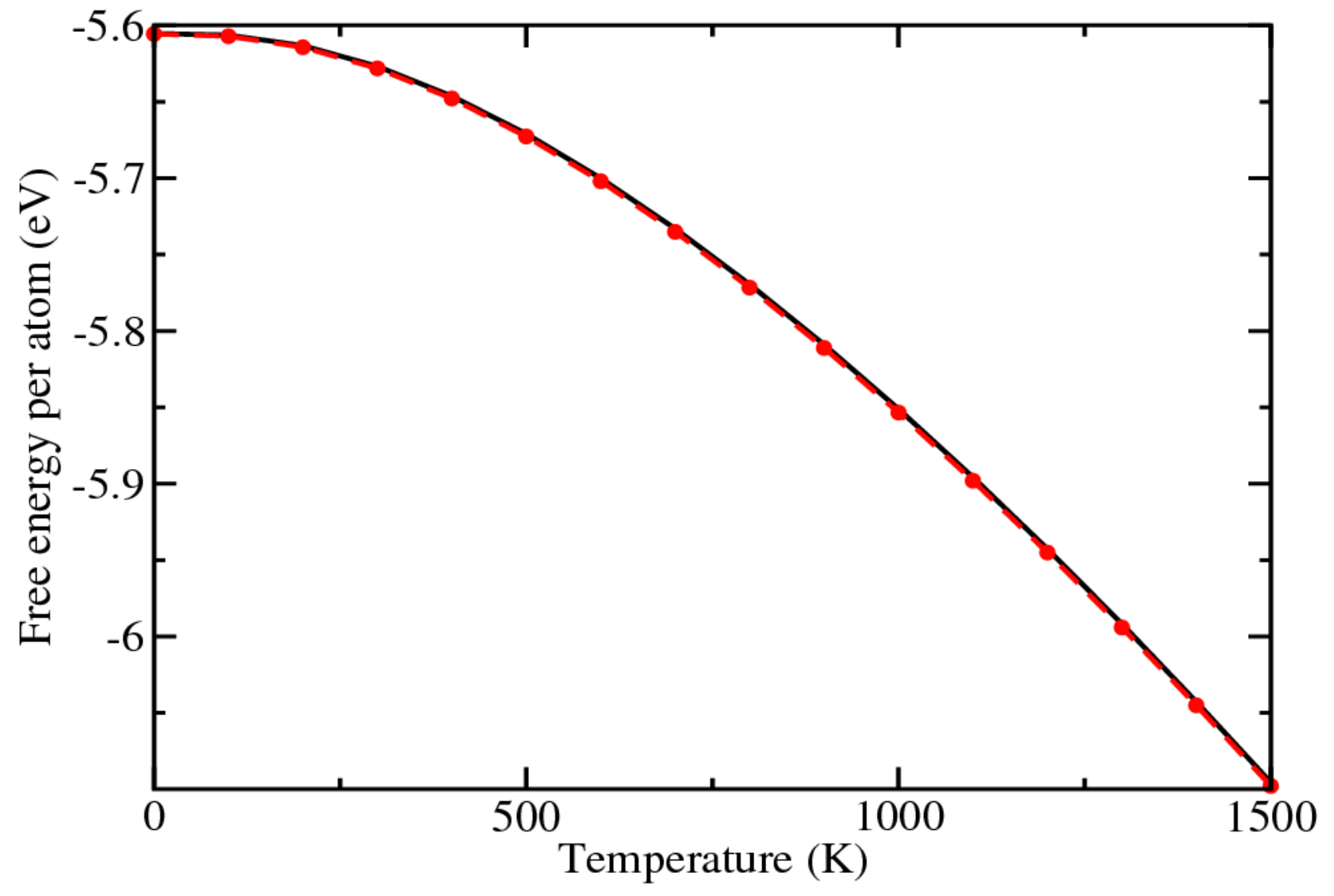}
\caption{Free energy for perfect BAs, calculated within the quasiharmonic approximation using the settings as described. The black line denotes a calculation based on a $4\times4\times4$ 512-atom supercell, the red dots and dashed line are based on a $2\times2\times2$ 64-atom supercell.}
\label{fig:QHA}
\end{figure}

While in the calculations in the paper the actual numerical results for the specific temperatures are used, for reference, we report a fit of the vibrational contribution here based on a fourth-order polynomial without linear term, which fits the actual DFT results over the temperature range within 0.01 eV per atom. As reference and for cross-validation, we report here results for solid As (relaxed minimum-energy $R\bar{3}m$ structure from MaterialsProject.org within a $6\times6\times2$ 432-atom supercell), B (relaxed minimum-energy $R\bar{3}m$ structure from MaterialsProject.org within a $3\times3\times1$ 324-atom supercell),  we find
\begin{equation}
    F_{\rm As,vib}(T) = 0.0252 - 7.7436\times 10^{-7}T^2 + 3.9500\times 10^{-10}T^3
    - 8.392\times10^{-14}T^4 \nonumber
\end{equation}
and
\begin{equation}
    F_{\rm B,vib}(T) = 0.1336 - 9.3379\times 10^{-8}T^2 - 6.4218\times 10^{-11}T^3
    + 2.4277\times10^{-14}T^4 \nonumber
\end{equation}
respectively. For BAs in zinc blende structure, we perform a convergence test with respect to supercell size. The results from a $4\times4\times4$ 512-atom supercell,
\begin{equation}
    F_{\rm BAs,vib,512}(T) = 2\left(0.0828  -3.1397\times 10^{-7}T^2 + 7.1954\times 10^{-11}T^3
    - 6.0088\times10^{-15}T^4\right) \nonumber
\end{equation}
are nearly identical to those from a $2\times2\times2$ 64-atom supercell,
\begin{equation}
   F_{\rm BAs,vib,64}(T) = 2\left(0.0820 - 3.2050\times 10^{-7}T^2 + 7.8671\times 10^{-11}T^3
    - 7.8746\times10^{-15}T^4\right) \nonumber
\end{equation}
as also can be seen in Fig.~\ref{fig:QHA}. 

\section{Thermodynamic Equilibrium of BA\lowercase{s} with Point Defects}
\label{Sec:SI_thermoequilibrium}

In the main text, for finite temperatures, the chemical potentials of As and B are determined from the requirement that thermal point defects maintain stoichiometry, i.e. that the system is closed and no atoms are exchanged with the environment when point defects are created. Within this approach, point defects are then coupled to stoichiometry, since they are the only way non-stoichiometry in a crystal such as B$_{1+\delta}$As$_{1-\delta}$ can be accommodated for $\delta \ne0$. The point defects that create the specific stoichiometry with the lowest energy penalty are called the ``constitutional defects.''

The concept of constitutional point defects and coupling point defects to the specific stoichiometry is based on the grand thermodynamic potential and is in use since about three decades, where first applications targeted intermetallics \cite{Mayer95,Schott97,hagen1998point} with later extension to insulating systems \cite{mishra2012native}. 
The ground state of an ordered binary compound that is off-stoichiometric, such as B$_{0.52}$As$_{0.48}$, is defined by the system with the lowest energy. Since it is off-stoichiometric, the ``perfect'' off-stoichiometric system needs to have constitutional point defects to accommodate the off-stoichiometry, for example antisites or vacancies. Assuming they don?t interact and thus no second phase precipitates, their formation energy with respect to the ``perfect'' non-stoichiometric system (which they define) has to be zero, otherwise they would not be the constitutional point defects. For perfectly stoichiometric crystals, {there are} no constitutional point defects, but only thermally activated ones according to their formation energies, whose concentration is typically many orders of magnitude lower than that of the constitutional point defects, since measurable off-stoichiometry is typically only created by defects with formation energies of at most a few tenths of an eV. 

In practice, the concentrations of all defects $X$ are expressed in the form of Eq.~(2) in the main manuscript, from which the concentrations of As and B atoms in BAs can be calculated by \cite{mishra2012native}
\begin{eqnarray}
{{N_{\rm As}}\over{V}} &&={N\over V} -  C_{V_{\rm As}} - C_{\rm B_{As}} + C_{\rm As_B}+C_{{\rm As}_i}, 
\label{As_conc}
\\
{{N_{\rm B}}\over{V}} &&={N\over V} -  C_{V_{\rm B}} - C_{\rm As_{B}} + C_{\rm B_{As}}+C_{{\rm B}_i}.  \label{B_conc}
\end{eqnarray}
For B$_{1+\delta}$As$_{1-\delta}$, the difference between the B and As concentrations needs to be $2\delta$, resulting in the equation 
\begin{equation}
C_{V_{As}}+C_{B_i}+C_{B_{As}}-C_{V_B}-C_{As_i}-C_{As_B}=2\delta{{N}\over{V}}.
\label{stoi_conc}
\end{equation}
Replacing the concentrations in Eq.~(\ref{stoi_conc}) with the corresponding expressions from Eq.~(2) with the free energies of formation following the form of Eq.~(1) creates a second equation in addition to $\mu_{\rm B} + \mu_{\rm As} = F({\rm BAs})$ as described in the main text that allows solving for the chemical potentials. Since the free energies depend on temperature and Eq.~(1) adds the Fermi level into the equation, the chemical potential $\mu(T,E_F,\delta)$ as a function of $T$ and $E_F$ for stoichiometry ($\delta=0$) or a given non-stoichiometry $\delta$ can be determined. 
\textcolor{black}{For a ternary system A$_m$B$_n$C$_l$, there will be four conditions following the notation in the main text
\begin{eqnarray}
    m\mu_{\rm A}+n\mu_{\rm B}+l\mu_{\rm C}&=&E(A_m B_n C_l)\\
    C_{\rm V_{A}}+C_{\rm B_i}+C_{\rm B_{A}}-C_{\rm V_B}-C_{\rm A_i}-C_{\rm A_B}&=&\delta_{\rm AB}{{N_{\rm A}+N_{\rm B}}\over{V}}\\
    C_{\rm V_{A}}+C_{\rm C_i}+C_{\rm C_{A}}-C_{\rm V_C}-C_{\rm A_i}-C_{\rm A_C}&=&\delta_{\rm AC}{{N_{\rm A}+N_{\rm C}}\over{V}}\\
    C_{\rm V_{B}}+C_{\rm C_i}+C_{\rm C_{B}}-C_{\rm V_C}-C_{\rm B_i}-C_{\rm B_C}&=&\delta_{\rm BC}{{N_{\rm B}+N_{\rm C}}\over{V}}
\end{eqnarray}
where only three in four equations are linearly independent, allowing us to solve the chemical potentials for A, B, and C respectively.}

\begin{figure}
\includegraphics[width=0.6\linewidth]{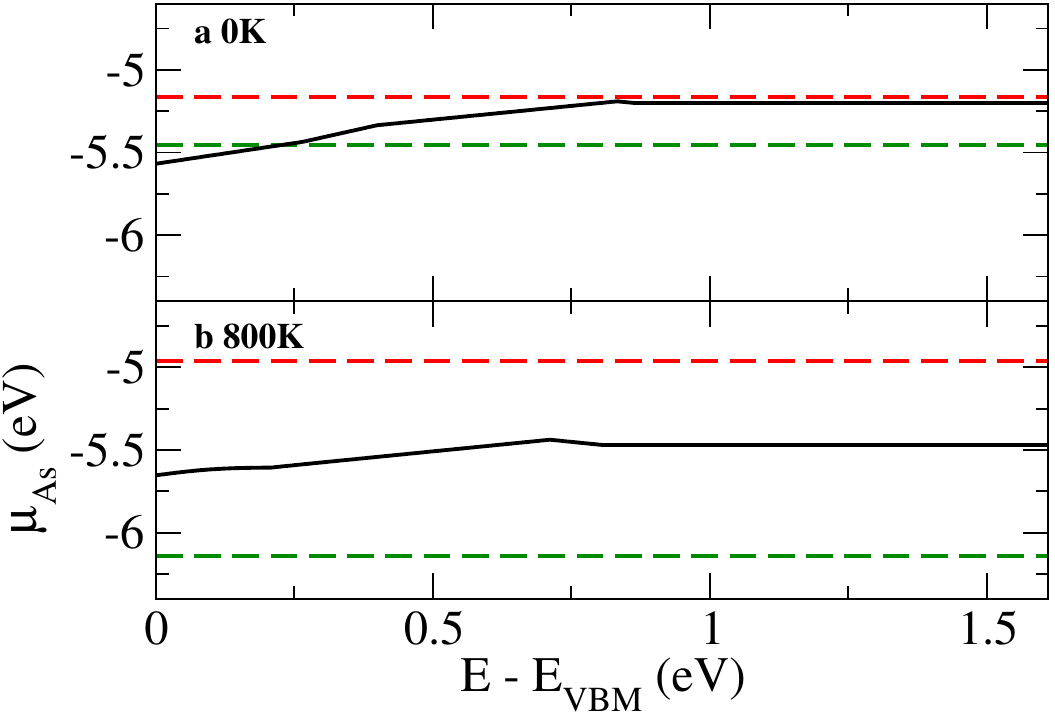}
\caption{ Chemical potential for As {in stoichiometric BAs} at (a) 0~K and (b) 800~K, calculated within the quasiharmonic approximation using the settings as described. The red and green dashed lines are the limits for As-rich and B-rich cases, respectively. Zero temperature does not include the zero-point energy contribution.}
\label{fig:muVsT}
\end{figure}
Next, we show that the chemical potentials generated this way fulfill the stability criterion and do not exceed the boundaries set by the B-rich and As-rich limits, i.e.\ the elemental chemical potentials of B and As. Figure~\ref{fig:muVsT} shows the As chemical potential as a function of Fermi level position as determined from the requirement of stoichiometry in comparison to the limits of As-rich and B-rich material, all calculated within QHA as described in Sec.~\ref{Sec:SI_QHA}. The figure shows that for elevated temperatures, the chemical potential is within the limits, indicating that the solution found from the stoichiometry requirement indeed fulfills the condition of thermodynamic stability for all Fermi level positions. At 0~K, the allowed range for the chemical potential of As ($\mu_\mathrm{As}$) is much narrower than that at higher temperatures, and for extreme-$p$ type conditions with Fermi levels $< 0.25$~eV, stoichiometric BAs is unstable. More restrictive than that, Zheng \textit{et al.}~\cite{zheng2018antisite} even argued that stoichiometric BAs at zero temperature is not a thermodynamically stable phase at all. To shine more light on this question, we examine the thermodynamic stability of stoichiometric BAs in Sec.~\ref{Sec:SI_convexhall}.
We further note that, elemental As has a sublimation point of $\sim 887$~K.
In fact, the upper limit of $\mu_{\rm As}$ should be the minimum value within all possible phases. 
Here we take the limit of $\mu_{\rm As}$ in its solid phase, calculated by the quasiharmonic approximation, because the chemical potential in the gas phase varies significantly upon the partial pressure and so on.

\section{transition from canonical to grand canonical at Off-stoichiometric limit}
\label{Sec:SI_delta}
\begin{figure}
    \centering
    \includegraphics[width=0.6\textwidth]{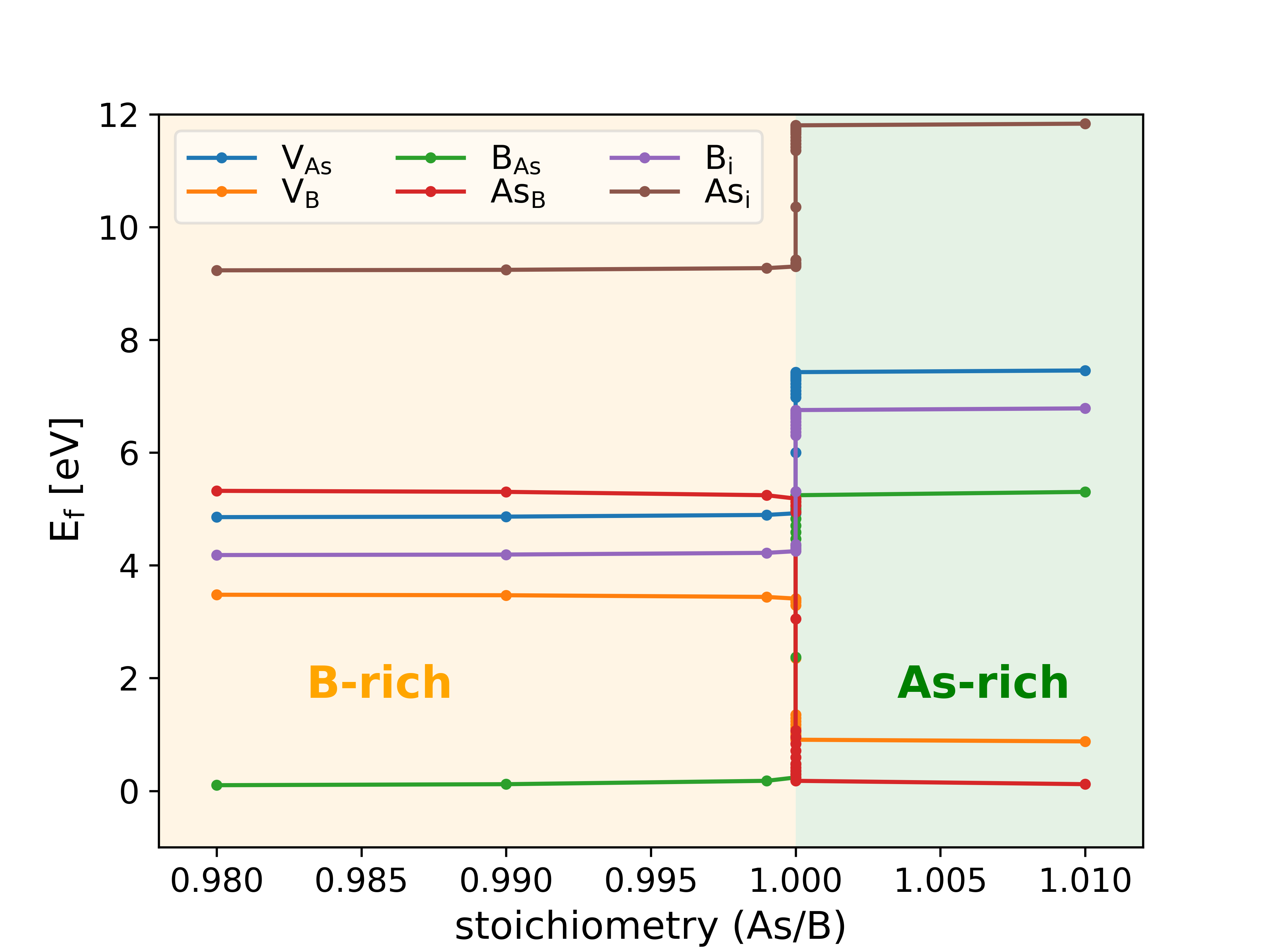}
    \caption{Defect formation energies at different crystal stoichiometry $0.98<{\rm As/B}<1.01$ evaluated at 300~K at $E_F\sim0.8$~eV, highlighting that the formation energy of the ``constitutional'' defect, i.e. B$_{\rm As}$ in the B-rich region and As$_{\rm B}$ in the As-rich region, approaches to zero quickly when the lattice is off-stoichiometric.}
    \label{fig:delta}
\end{figure}

\textcolor{black}{As discussed in \emph{Sec.~IIA Theoretical framework} in the main text, our canonical approach allows the chemical potentials, thus the defect formation energies, in a compound to be determined from a specific lattice stoichiometry B$_{1+\delta}$As$_{1-\delta}$. 
Here we show in Fig.~\ref{fig:delta} that, in BAs, the self-consistent solution for the chemical potentials do not vary upon the atomic ratio of As/B and the formation energies of the lowest energy defect approaches zero when the off-stoichiometry $\delta$ is on the order of $\pm10^{-6}$.
These lowest formation energy defects are called ``constitutional defects''.
Therefore, for an experimentally detectable composition, the off-stoichiometry limit thus aligns with the grand canonical approach practically. 
However, instead of being equal to the elemental limit, the chemical potentials are determined by the condition that the constitutional defect has zero formation energy while others have positive values.}

\section{Phase Stability of Cubic Stoichiometric BA\lowercase{s}  }
\label{Sec:SI_convexhall}
In this section we discuss the phase stability of BAs with respect to its left and right neighbors in the phase diagram, elemental B and B$_6$As, by constructing the convex hull at different temperatures from free energies of formation. To consider the effect of thermal expansion, the temperature-dependent lattice constants of the four phases, i.e.\ elemental B~\cite{tsagareishvili1986thermal}, elemental As~\cite{fischer1978debye}, pure BAs~\cite{Chen2019}, and pure B$_6$As~\cite{Whiteley2013} are taken from experiments to include thermal expansion as lowest-order anharmonic effect into the QHA framework as described in Sec.~\ref{Sec:SI_QHA}.
\begin{figure}[!ht]
\graphicspath{{./figures/}}
\includegraphics[width=0.6\linewidth]{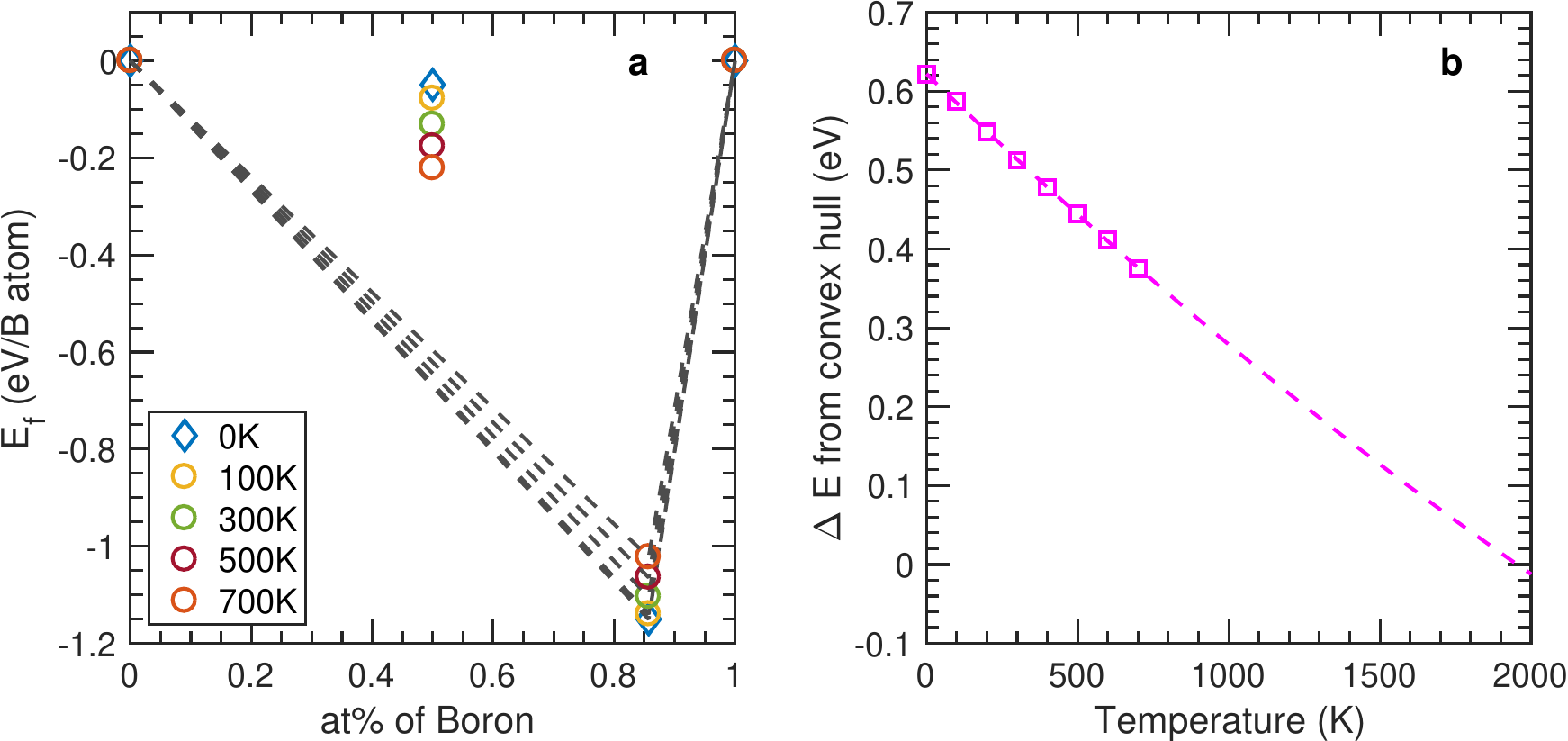}
\caption{\textbf{a} Convex hull plot of the formation energy (open diamond) and free energies of formation (open circles) E$_f$ of the B-As phases (BAs and B$_6$As) in eV/B~atom calculated by quasi-harmonic approximation. \textbf{b} The energy difference between the BAs phase and the convex hull in eV/B~atom as a function of temperature, approaching zero at elevated temperatures. Dashed line shows a second-order polynomial fit.}
\label{fig:ConvHull} 
\end{figure}

Figure~\ref{fig:ConvHull} shows the convex hull plot \cite{Hautier2014} for the formation energies of the BAs and B$_6$As phases. The entropy contribution increases the stability of the BAs phase while decreasing that of the B$_6$As phase. By calculating the energy difference between the free energies of formation of BAs and the convex hull, the beneficial effect of elevated temperature can be clearly shown (Fig.~\ref{fig:ConvHull}b). Only energies up to 700~K are shown because As has a sublimation point $\sim 887$~K. 
Extrapolating from the values up to 700 K with a second-order polynomial fit estimates that the BAs phase will be located on the convex hull at temperatures around 1900~K. This estimate does not consider the excess As:B ratio used in synthesis, which may modify the thermodynamic equilibrium to lower temperatures and enable growth of stoichiometric BAs. This is consistent with the fact that at elevated temperatures, the chemical potential range allowed for stoichiometric BAs is much wider than that at 0~K, as shown in Fig.~\ref{fig:muVsT}.

\providecommand{\noopsort}[1]{}\providecommand{\singleletter}[1]{#1}%
%

\end{bibunit}
\end{document}